# EBSD and TKD analyses using inverted contrast Kikuchi diffraction patterns and alternative measurement geometries

Grzegorz Cios[a,*], Aimo Winkelmann[a], Gert Nolze[b,c], Tomasz Tokarski[a], Benedykt R. Jany[d], Piotr Bała[e,a]

[a]*Academic Centre for Materials and Nanotechnology, AGH University of Krakow, al. A. Mickiewicza 30, 30-059 Krakow, Poland*
[b]*Federal Institute for Materials, Research and Testing (BAM), Unter den Eichen 87, 12205 Berlin, Germany*
[c]*TU Bergakademie Freiberg, Institute for Mineralogy, Brennhausgasse 14, 09596 Freiberg, Germany*
[d]*Marian Smoluchowski Institute of Physics, Faculty of Physics, Astronomy and Applied Computer Science, Jagiellonian University, 30348 Krakow, Poland*
[e]*Faculty of Metals and Industrial Computer Science, AGH University of Krakow, al. A. Mickiewicza 30, 30-059 Krakow, Poland*

**Abstract**

Electron backscatter diffraction (EBSD) patterns can exhibit Kikuchi bands with inverted contrast due to anomalous absorption. This can be observed, for example, on samples with nanoscale topography, in case of a low tilt backscattering geometry, or for transmission Kikuchi diffraction (TKD) on thicker samples. Three examples are discussed where contrast-inverted physics-based simulated master patterns have been applied to find the correct crystal orientation. As first EBSD example, self-assembled gold nanostructures made of Au fcc and Au hcp phases on single-crystal germanium were investigated. Gold covered about 12% of the mapped area, with only two thirds being successfully interpreted using standard Hough-based indexing. The remaining third was solved by brute force indexing using a contrast-inverted master pattern. The second EBSD example deals with maps collected at a non-tilted surface instead of the commonly used 70˚ tilted one. As TKD example, a jet-polished foil made of duplex stainless steel 2205 was examined. The thin part close to the hole edge producing normal-contrast patterns were standard indexed. The areas of the foil that become thicker with increasing distance from the edge of the hole produce contrast-inverted patterns. They covered three times the evaluable area and were successfully processed using the contrast-inverted master pattern. In the last example, inverted patterns collected at a non-tiled sample were mathematically inverted to normal contrast, and Hough/Radon-based indexing was successfully applied.

*Keywords:* EBSD, TKD, contrast inversion, topography, Kikuchi diffraction,

## 1. Introduction

In scanning electron microscopy (SEM), the technique of electron backscatter diffraction (EBSD) is a common tool for a crystallographic microstructural characterization of materials. Analyses of local orientation, misorientation or phase distribution are still the most frequent tasks of EBSD. The standard procedures for Kikuchi diffraction pattern indexing, which are applied in commercial EBSD systems, are based on the assumption that, relative to the local background, the Kikuchi bands exhibit a slightly increased intensity within a width of approximately twice the Bragg angle from the corresponding lattice plane traces.

It is well-known that EBSD Kikuchi bands can also exhibit lower intensity than the background signal under specific measurement conditions [1]. This is the case when the influence of inelastic scattering processes on the outgoing diffraction processes is increased [2]. Even within a single Kikuchi pattern, the intensity distribution can change continuously from a situation where the Kikuchi bands have higher intensity than the local background to a case where the Kikuchi band shows lower intensity than the background ("dark bands"). The role of inelastic scattering indicates that dark Kikuchi bands should be observed in all cases where the scattered electrons have to go trough an increased thickness of material. This includes, for example, thicker TKD samples, or when electrons are transmitted through topographic features. Dark bands are also observed in the conventional backreflection geometry when the effective Kikuchi diffraction source depth is increased due to a steeper angle of incidence of the primary SEM beam in combination with low take-off angles.

The contrast inversion effects can be understood semi-quantitatively by including anomalous absorption in a dynamical electron diffraction model [2]. Concerning the extension of the application cases of EBSD and TKD, the assumption of Kikuchi bands with only increased intensity is a limitation of the conventional Hough/Radon transform-based standard orientation determination, as well as for pattern matching techniques comparing simulated patterns with experimental ones but neglecting the specific effects of anomalous absorption.

*Corresponding author
Email address:* Ciosu@agh.edu.pl (Grzegorz Cios)

The present paper presents case studies where Kikuchi diffraction patterns with dark bands limit the solution rate in orientation maps. We demonstrate that EBSD and TKD can be applied in an extended range of experimental conditions compared to conventional measurements using normal contrast Kikuchi diffraction.

## 2. Materials and Methods

### 2.1. Samples, preparation and equipment used

The Au nanoislands on Ge were prepared in an ultrahigh vacuum (UHV) molecular beam epitaxy (MBE) system. The surfaces of the Ge(0 0 1) substrate were cleaned by cycles of low-energy ion beam bombardment and annealing in order to achieve atomically flat terraces. Deposition of 30 monolayers (ML) of Au by the MBE at 673K resulted in the formation of Au nanoislands on the sample surface made of Au fcc and Au hcp phases[3].

The Ni sample for EBSD investigations was cut from a technically pure Ni rod and prepared by a standard metallographic procedure, i.e. grinding on SiC papers, polishing on diamond suspensions 1 and 3µm and vibration polishing on 50 nm colloidal silica for 5 hours.

As TKD samples 2205 grade steel as well as single grain transformer steel were ground down to foils of $\leq 100$ µm thickness and electropolished in a twin-jet device at 25 V. As electrolyte served a 10% perchloric acid solution in glacial acetic acid.

The EBSD setup applied in this study consists of a field emission gun SEM Versa 3D (FEI), which was equipped with Symmetry S2 EBSD detector (Oxford Instruments Nanoanalysis). The sample preparation was carried out using a vibration polisher QPOL Vibro (QATM), and a Tenupol-5 (Struers) electro-polisher. The MBE system was supplied by Prevac.

The applied data acquisition parameters are given in Table 1.

Table 1: Data acquisition parameters

| sample | applied technique | surface tilt | $E_o$ [keV] | beam current [nA] | step size [nm] | camera mode | speed [patterns/s] |
|---|---|---|---|---|---|---|---|
| Au/Ge | EBSD | 70° | 15 | 10 | 30 | Speed 1 | 280 |
| Ni | EBSD | 0° | 30 | ≈ 30 | 200 | Speed 2 | 95* |
| 2205 duplex | TKD | −20° | 30 | ≈ 30 | 100 | Speed 2 | 745 |
| transformer steel | TKD | −20° | 30 | ≈ 30 | 500 | Speed 1 | 600 |

\* 4 averaged frames for noise reduction

### 2.2. Orientation determination

For further analysis, all diffraction patterns were stored in a single file using AZtec v. 6.1 (Oxford Instruments Nanoanalysis) and subsequently transferred into a HDF5-compatible [4] file format of type *.h5oina [5].

For pattern matching the commercial software MapSweeper (AZtec Crystal v. 3.1) was used. The required physics-based master pattern has been simulated in AZtec Crystal as well. The dynamical simulations are based on the same approach described in [6]. For the interpretation of inverted-contrast diffraction patterns, the master pattern of the phase was intensity-inverted and then stored as a new master pattern, i.e., each crystallographic phase is described by two apparently different phases. For the inverted master pattern, no special simulation was performed. Intensities from the simulated normal contrast master pattern were inverted. One is presented by the normal intensity master pattern whereas the second one shows the inverted contrast master pattern (see Fig. 1).

During brute force analysis, each experimental pattern was compared with real-time generated patterns from both normal contrast and inverted contrast master patterns. The orientation space was parsed with an initial step size of 2°. Using the normalized cross-correlation coefficient $R$ as quality criterion, the final orientation results from a further refinement using the Nelder-Mead approach [7]. The minimum acceptable R-values for indexing were set to $R > 0.15$.

## 3. Results and Discussion

### 3.1. EBSD from samples with surface nanostructures

The distribution from Au nanoislands on Ge(0 0 1) is shown in the phase distribution map in Fig. 2. During EBSD mapping the sample was tilted by 70° so that the Au islands have a height of ≤ 20 nm.

The shape of the Au nanoislands in combination with the tilt of the Ge sample is responsible for a continuously varying angle $\chi$ between primary beam and local surface normal. Due to the scanning in this case from bottom left to top right for EBSD measurements, contrast-inverted Kikuchi patterns appear for $\chi \ll 70°$. The phase distribution map overlaying the band contrast is shown in Fig.2. Due to the bottom-up scanning, non-indexed tops of the nanoislands appear at their lower ends. Because of this topography almost 30% of all Au patterns remain non-indexed. The investigations also show that gold occurs in the present microstructure as two closed-packed modifications, the face-centered cubic (fcc) and the hexagonal closed-packed (hcp) phase.

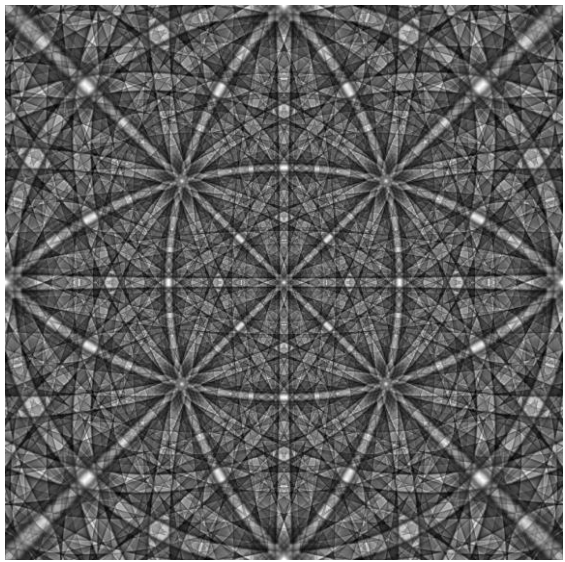

(a) normal intensity master pattern

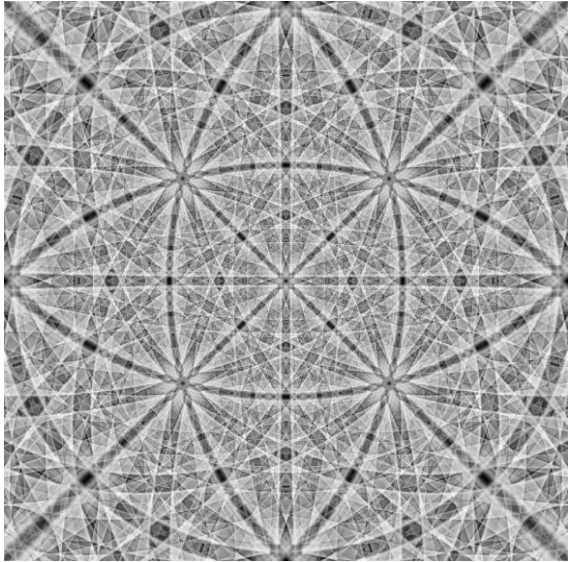

(b) inverted intensity master pattern

Figure 1: Master pattern for the interaction of BCC iron with 20 keV electrons displayed as stereographic projection. a) shows the normal intensity, whereas b) reflects the inverted intensity.

Two examples of experimental and template-matched patterns are presented in Fig.3 a) and b). Both are from the same nanoisland, see A and B in Fig.2. The match of the normal contrast fcc-Au pattern in Fig.3(a) results in a high normalized cross-correlation coefficient ($R = 0.7200$).

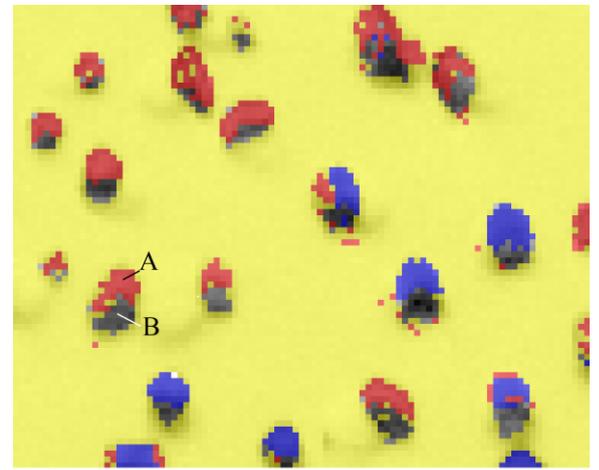

Figure 2: Band contrast overlaid by the phase distribution as received from EBSD software: 87% Ge (yellow), 6% fcc-Au (red), 3% hcp-Au (blue). A and B indicate the location of patterns discussed in Fig. 3. Scale bar length: 1 μm.

However, the inverted contrast pattern gives a much lower $R = 0.4506$, which is due to shadowing in the upper region and the generally higher noise in the pattern.

The applied pattern matching for all phases, including their inverted contrast signals, is shown as $R$ and phase distribution map in Fig. 4. Normal contrast and inverted contrast patterns are shown as separate colors.

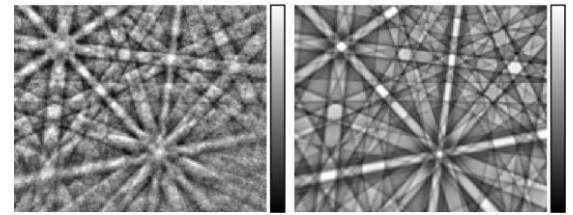

(a) Experimental pattern from A and simulation, $R = 0.7200$

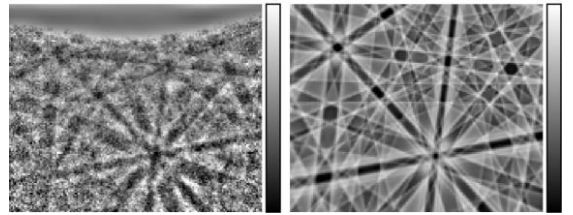

(b) Experimental pattern from B and simulation, $R = 0.4506$

Figure 3: Comparison of experimental patterns (left) with the according simulations (right) for two example position in a fcc-Au nanoisland, see Fig.2). $R$ is the normalized cross-correlation coefficient.

The very low correlation separating regions of normal and inverted contrast patterns in Fig. 4 can be explained by their overlay. The upper half is thus dominated by the normal signal and the lower half by the contrast-inverted signal. In this case, the cross-correlation for both master patters is poor.

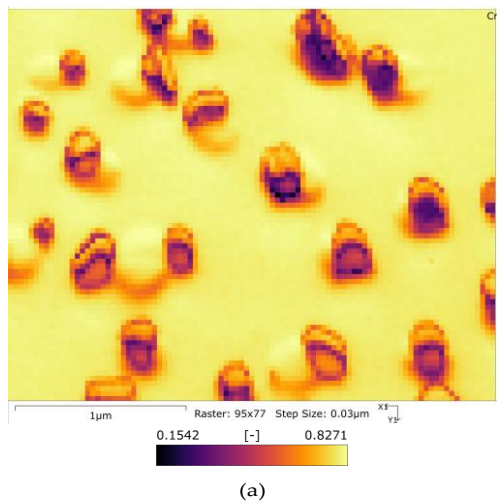

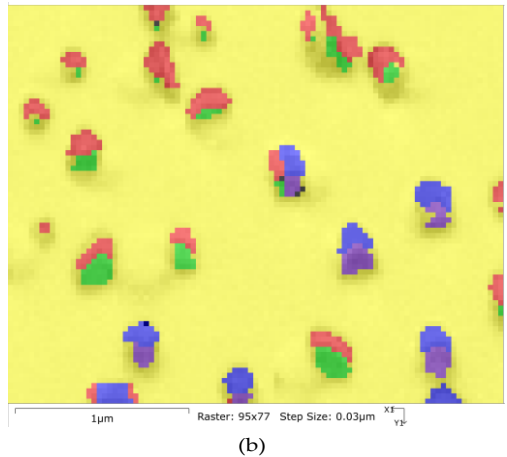

Figure 4: Au nanoislands grown on Ge(0 0 1). (a) shows the normalized cross-correlation coefficient after pattern matching with the normal as well as contrast-inverted master patterns. (b) displays the band contrast overlaid by the identified phase: yellow (Ge, 90%), red (fcc-Au, 4%), blue (hcp-Au, 2.%). The green (fcc-Au, 2%)and purple (hcp-Au, 1%) colored pixels result from contrast inverted patterns.

### 3.2. Large area TKD from thick samples

TKD is usually limited to a relatively small size of the analyzed area compared to typical EBSD measurements. The reason is that the preparation of large-area thin samples with constant thickness is rarely possible [8]. For metallic materials, the application of twin-jet electropolishing is recommended. Unfortunately, the thickness of such samples increases continuously with the distance from the hole created during polishing. A larger sample thickness does not only reduce the spatial resolution, but there is also a gradual contrast reversal of the TKD patterns. This limits the field of investigation further, as the standard algorithms do not work reliably with contrast-reversed patterns. The phenomenon of contrast inversion in TKD patterns is demonstrated in Fig. 5 on single crystal transformer steel sample thin foil. Pattern 1 shows a high-quality normal contrast. The contrast of pattern 4, taken at a point about 50 µm from the hole from electropolishing, is already partially inverted, best seen by the vertical narrow band in the middle of the pattern. The bands are of higher intensity than the background in the upper half but become lower than the background in the lower half. Pattern 5 seems to be completely inverted, however, in comparison to pattern 6 the contrast appears to be higher due to the increasing noise. The patterns in Fig. 5 demonstrate that an additional indexing of contrast-inverted TKD patterns may increase the size of an analyzable area by a factor of 3.

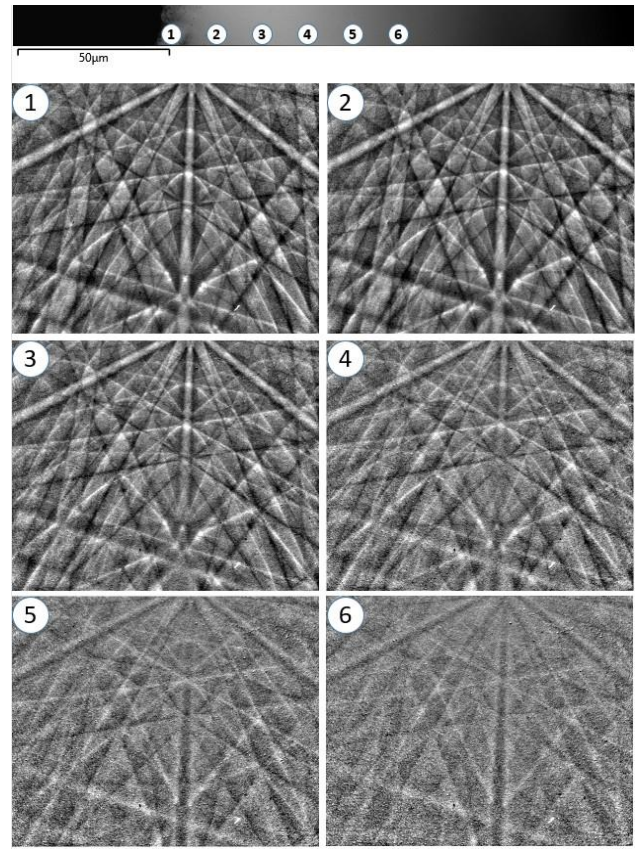

Figure 5: TKD patterns collected from single crystal thin foil (transformer steel). The according FSD image on the top is used to mark positions where the pattern 1–6 have been captured. The area on the left side of spot 1 is the hole from electropolishing.

Fig. 6 demonstrates the significantly enlarged analysis area through the additional indexing of inverted contrast patterns. In the background, the microstructure captured by forescattered electron detector is displayed. The insets a) – c) represent typical TKD patterns for three different thicknesses and the corresponding simulations. Insets d) - f) show inverse pole figure (IPF) maps for the sample reference direction Z, with a size of 20×50µm areas overlaying the microstructure displayed in the background. f) was indexed using normal contrast master patterns, whereas d), and e) showed better matches using inverted contrast master patterns. The respective normalized cross-correlation maps are shown in the images g) – i). Caused by the larger electron beam spread due to a increasing thickness, the grain boundaries in h) and g) appear significantly broader. This indicates that with increasing thickness, the spatial resolution decreases, although the corresponding IPF maps e) and d) are not affected at all. This underlines that TKD analysis in sample areas thicker than usually recommended still provides a sufficient spatial resolution.

The example in Fig. 6 demonstrates that using inverted-contrast master patterns enables the analysis of bigger areas with various thickness also by TKD. Fig. 6a 1 to 4 show pairs of experimental (left) and simulated (right) patterns extracted from the TKD map whose band contrast is shown in Fig. 6b. Patterns are numerated from the thinnest sample to thickest 1 to 4 respectively. Notable is pattern 2 for which R has almost the same value for normal and inverted contrast. Fig. 6c) to e) show phase R and IPF maps respectively where dictionary indexing was performed using BCC and FCC normal and inverted contrast master patterns in maps f) to h) maps were dictionary indexed using normal contrast master patterns and i) to k) using inverted contrast master patterns. Comparing phase maps (c, f, i) and IPF maps (e, h, k) it is clear for indexing using normal and inverted master patterns qualitatively smallest number of misindexed points is found. Panel c) in Fig. 6 shows the area of analysis, which has been increased approximately three times with the possibility to index the inverted patterns.

*3.3. Low tilting angle EBSD*

The commonly recommended practice to use high-tilted sample for EBSD investigations has its origin in the increase of the backscatter coefficient with the tilt angle $\phi$. In [9] two different approximations are given which are either related to the mean atomic number $Z$:

$$\eta_\phi = f(Z, \phi) = (1 + \cos\phi)^{-9/\sqrt{Z}} \qquad (1)$$

or directly to the backscatter coefficient of a non-tilted sample $\eta_0$:

$$\eta_\phi = f(\eta_0, \phi) = 0.89 \left(\frac{\eta_0}{0.89}\right)^{\cos\phi} \qquad (2)$$

where $\eta_0$ is derived from $Z$ by

$$\eta_0 = -0.0254 + 0.016Z - 0.000186Z^2 + 8.3 \times 10^{-7}Z^3$$

given in [9] as well.

Fig. 7 demonstrates that for $\phi = 70°$ equations (1) and (2) deliver similar results. The gain $\eta_{70}/\eta_0$ is practically identical. Fig. 7 mainly indicates that the signal amplification by $\eta_\phi$ decreases with increasing $\eta_0$ (or $Z$). The gain can be up to 10 times higher for materials with low scattering than for samples tilted by 70°. The factor, on the other hand, decreases to less than double for heavy phases. This explains, why samples containing light as well as heavy phases can be analyzed simultaneously with EBSD. $\eta_{70}$ can only vary between $0.25 < \eta_{70} < 0.72$, see Fig. 8 (in contrast to a factor 100 for non-tilted surfaces). Optimizing the dwell time for the heaviest phase means that the signal for the weakest possible phase only uses about 30% of the camera dynamic, which is, in many cases, sufficient. Considering that a large number of phases in the untilted state have BSE yields between $0.15 < \eta_0 < 0.4$, the effective BSE yields $\eta_{70}$ of the affected phases after 70° tilt are even less different.

However, $\eta$ describes the total amount of all BSE described by a directional distribution that depends on $Z$ as well as on the sample tilt $\phi$, cf. e.g. [9]. Since the projection center (PC) defines the position of the point source of the signal with respect to the coordinate system of the detector screen, and the the detector screen captures only a certain sector of this distribution, the estimates of $\eta$ for highly tilted samples should only be interpreted in a qualitative sense, because the measured integrated intensities on the EBSD screen will depend on the actual measurement geometry (PC).

Fig. 8 also illustrates the expected increase in $\eta_\phi$ for increasing $\phi$.

However, the increase of $\eta$ by a high tilt of the surface brings several disadvantages. The assumed circular beam degenerates into a clear elliptical cross-section so that the resolution in y-direction is worse than in x, see Fig. 9. In order to map the surface in an apparently equidistant manner, the surface is scanned narrower along y by the factor of the tilt correction, which is equal to $1/\cos\phi$.

However, there are further purely geometrical drawbacks. Any sample topography, mainly caused during preparation, reduces the comparability of surface images taken vertically or at a high tilt angle. In addition, for high-tilted surfaces imaged at low magnifications, the focus must be noticeably tracked. Moreover, a typically rectangular measurement field ideally becomes trapezoidal. Superimposed on this is a magnification-independent rhombus-shaped distortion of the measuring field which increases with the tilt angle. This is only excluded if the surface is exactly perpendicular to the primary beam at 0° tilt, which is very rarely the case in practice [10].

The surface tilt of exactly 70°, which is often suggested for EBSD, is not evident by any special properties from any of the above diagrams in Fig. 7-9, but it is due to practical historical reasons when a Si(0 0 1) sample was used for system calibration during installation since $\angle(\langle 0\,0\,1\rangle, \langle 1\,1\,4\rangle) = 19.47° \approx 90° - 70°$.

Fig. 7 and 8 indicate that in principle any sample tilt angle should be useable for EBSD examinations as long as the applied detector is sensitive enough. For example, horizontally aligned sample surfaces would eliminate all the disadvantages listed above for high tilt angles. In line with this expectation, in [11], a new design of a horizontally aligned direct electron detector has been introduced parallel to the sample surface.

As an alternative to a setup with both sample surface and detector aligned horizontally, a standard EBSD detector can be moved in from the side, resulting in a tilt-free sample and a conventional, vertical phosphor screen, see Fig. 10. The vertical screen collects, however, a part of the BSE signal that is emitted at grazing angles to sam-ple surface, which causes a contrast reversion in the lower part of the detector as described in [1]. Moreover, to prevent shadowing on the screen, the working distance needs to be increased so that the sample surface is sufficiently underneath the lower edge of the phosphor screen, see the sample/detector setup sketched in Fig. 10.

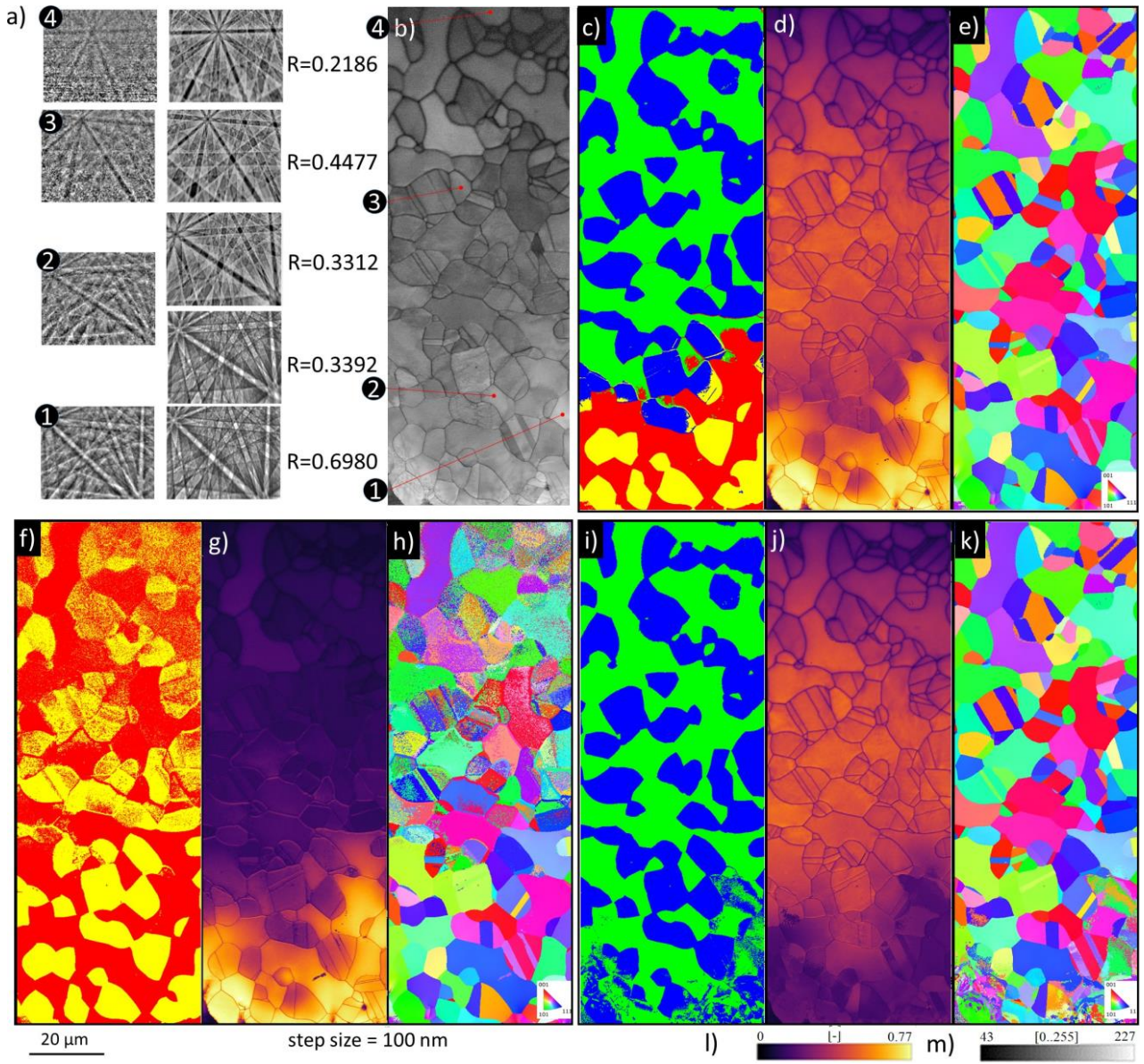

Figure 6: TKD maps from 2205 steel thin foil, a) experimental and simulated patterns, b) band contrast map with points where the patterns shown in a) were taken, c) - e) result of dynamic template matching with normal and inverted contrast master patterns, phase map, NCC, and IPF map respectively, f) - h) result of dynamic template matching with normal contrast master patterns, phase map, NCC, and IPF map respectively, i) - k) result of dynamic template matching with inverted contrast master patterns, phase map, NCC, and IPF map respectively, l) color bar for NCC maps, m) band color bar. For phase maps red is normal contrast BCC, yellow is normal contrast FCC, green is inverted contrast BCC and blue is inverted contrast FCC.

Fig. 11 shows that very low tilt angles close to 0° lead to the observation of BKD patterns that are completely contrast-inverted. They can be analyzed via pattern matching as described for the TKD examples above. For the Ni patterns in Fig. 11, tilt angles in between, i.e. for $10° < \phi > 60°$, lead to BKD signals showing normal contrasts in the upper part and reversed contrasts in the lower part; (Fig. 11 c) + d). Since the size of the contrast-reversed part in a BKD pattern depends again on the projection center, in addition to the tilt angle, $PC_y$ is given as a main component in Fig. 11 for each pattern. Using the simple approach of a single master pattern, so far, such mixed-contrast BKD patterns cannot be successfully interpreted either with the standard Hough techniques or with dictionary indexing using a master pattern or its inverse. However, adaptations of the software inverting only fixed parts of the simulated patterns, could be used to solve also such mixed contrast images.

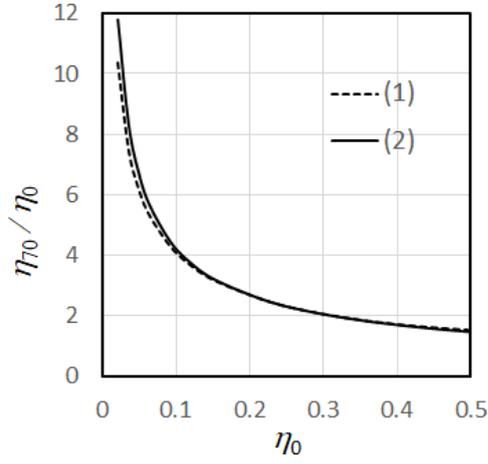

Figure 7: Impact of a 70° tilt on the increase of the backscattered signal.

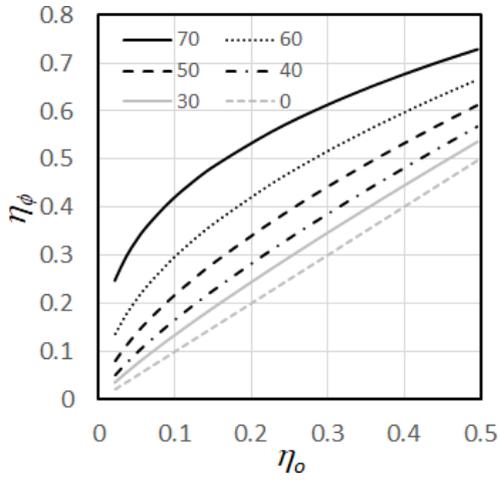

Figure 8: Impact of the tilt $\phi$ on the increase of the BSE yield $\eta_\phi$. $\eta_o^{Fe} = 0.279$, $\eta_o^{Ni} = 0.295$, $\eta_o^{Ge} = 0.323$, $\eta_o^{Au} = 0.487$.

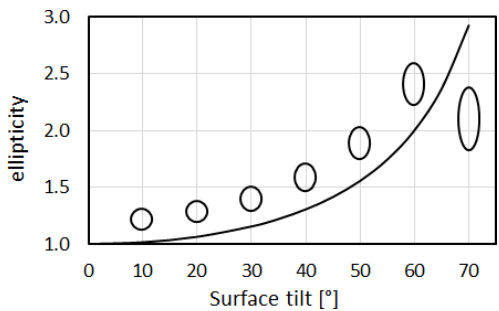

Figure 9: Ellipticity of the projected interaction volumeshape caused by surface tilt. The curve shows that at 70°, the beam spot along y is about three times longer than along x. To create an equidistant map of the surface, the primary beam is only deflected by about 1/3 along y, compared to the angle used for x.

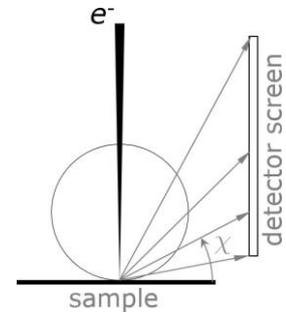

Figure 10: EBSD setup for investigation of non-tilted samples. The circle illustrates the typical spherical, directional intensity distribution of $\eta_o \propto \sin \chi$ for horizontal samples [9]. The characteristic BKD pattern (Fig. 11(e)), however, is only a weak modulated signal formed on the distribution of $\eta$. Its projection on a flat screen shows high gnomonic distortions.

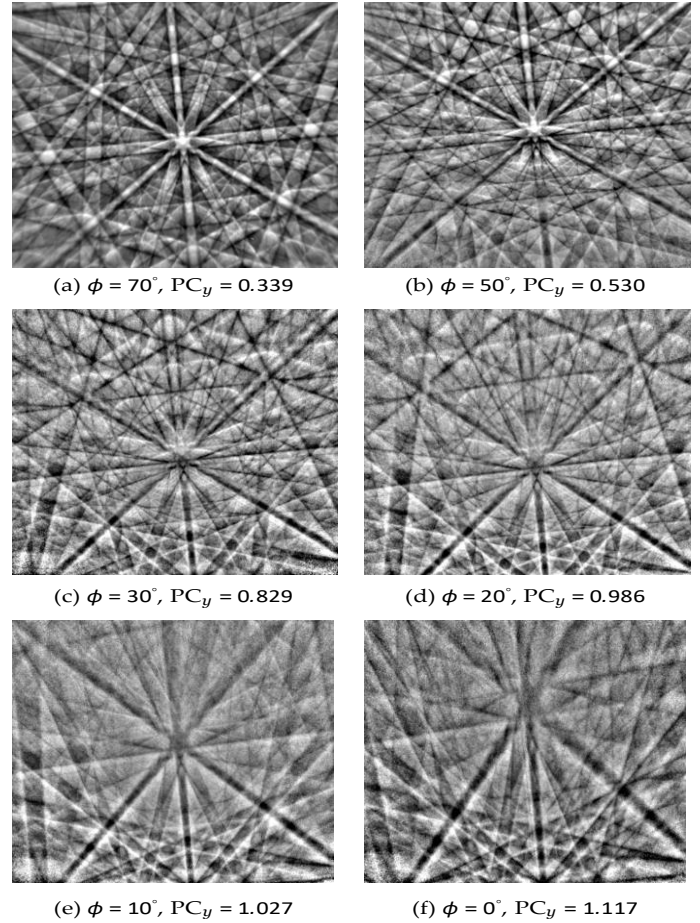

(a) $\phi = 70°$, $PC_y = 0.339$
(b) $\phi = 50°$, $PC_y = 0.530$
(c) $\phi = 30°$, $PC_y = 0.829$
(d) $\phi = 20°$, $PC_y = 0.986$
(e) $\phi = 10°$, $PC_y = 1.027$
(f) $\phi = 0°$, $PC_y = 1.117$

Figure 11: Change of the contrast in Ni patterns ($E_o$ = 30kV) with decreasing surface tilt $\phi$ and simultaneously increasing working distance WD. WD has not been specially optimized so that the individual $PC_y$ only describe the trend.

Also, slightly more tilted samples would hardly be affected by disadvantages resulting from tilt-like resolution differences (Fig. 9), projective distortions, or defocusing. In addition, on the one hand, $\eta_\phi$ as the overall BSE signal, but on the other hand, the visibility of the Kikuchi bands (BKD signal) would increase (Fig. 11), so that a slight increase of the tilt angle would benefit the measurement speed but also the resulting orientation accuracy.

An example of a Ni microstructure investigated on a sample at 0˚ tilt is shown in Fig.12. Here, a standard detector geometry is used, whereas in [12] where 0-degree tilt EBSD was demonstrated with the use of contrast inverted patterns, EBSD detector was pointing down from the flange where EDS detectors are typically mounted (therefore the patterns presented in [12] are not as heavily distorted as 11f). The EBSD dataset has been collected in about 80 min with a speed of 95 patterns/sec overlaying 4 frames in a single pattern. Using AztecCrystal MapSweeper, the post-processed brute force dictionary indexing and orientation refinement took about another 210 min[1] with a minimum acceptable correlation coefficient of $R_{min} = 0.15$. The IPF map for reference direction Z in Fig.12 displays the raw dictionary indexed and orientation refined data and is not cleaned. The hit rate was 99.8 %. In Appendix B similar dataset was indexed using Hough/Radon transform, showing the potential of implementation of the pattern invertion in commercial software.

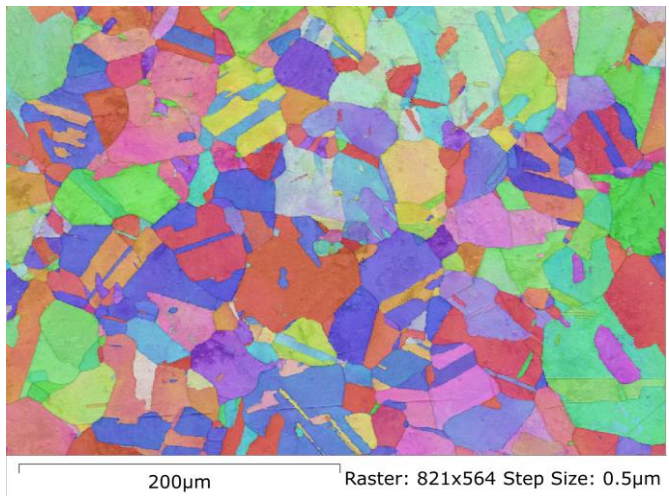

Figure 12: EBSD evaluation at a 0° tilted Ni-sample using 30 keV electrons.

## 4. Summary

In this study, three examples were used to demonstrate the successful application of inverted contrast Kikuchi patterns in EBSD pattern matching techniques for orientation interpretation and phase identification. The application areas include:

- conventional EBSD from nanostructures with pronounced surface topography

---
[1]On an Intel Xeon W-2145 SSD workstation, 64 GB of RAM, NVIDIA Quadro P400 GPU

- transmission Kikuchi diffraction (TKD) on samples with increased thickness
- low spatial distortion and improved spatial resolution EBSD measurements from low-tilt samples

The inverted image contrasts that occur under such acquisition conditions so far were very challenging to evaluate using standard evaluation methods, leading to a critical reduction in the indexing success. Using inverted master patterns in a pattern matching approach, it has been possible to assign these difficult diffraction patterns to unique orientations. The Table 2 summarizes the improvement in the hit rate when using an inverted master pattern (MP) or a combination of normal and inverted MP. The statistical significance of the obtained results is discussed in the Appendix A. As a proof of concept, it was also shown that the mathematical inversion of inverted contrast patterns makes these indexable with the Hough/Radon approach. This opens an opportunity for commercial EBSD software to test every non-indexed pattern to see if it is indexable after the inversion of its contrast.

Table 2: Hit rate for the investigated example microstructures using standard Hough-based indexing (in Oxford Instruments Aztec) and Pattern Matching (PM) in Oxford Instruments AztecCrystal.

| sample | applied indexing technique | |
|---|---|---|
| | standard Hough-based | normal and/or inverted MP PM |
| Au/Ge(0 0 1) | 96% | 100% |
| Ni (0˚ tilt) | 33% | 99.8% |
| 2205 (TKD) | 43% | 99.4% |

**Acknowledgments**

This work was supported by the Polish National Science Centre (NCN), grant no. 2020/37/B/ST5/03669.

This research was supported in part by the Excellence Initiative - Research University Program at the Jagiellonian University in Krakow.

The research results presented in this paper have been developed with the use of equipment financed from the funds of the "Excellence Initiative - Research University" program at the AGH University of Krakow. Research project partly supported by program „Excellence initiative – research university" for the AGH University.

**Data availability**

The raw data required to reproduce the above findings are available to download from Zenodo https://doi.org/10.5281/zenodo.10808974.

## Appendix A

The statistical significance of the obtained results was calculated for the Pearson correlation coefficient R using the code for EBSD Dictionary Based Indexing (DI) Confidence Maps available at [13]. In addition, the statistical uncertainty resulting from the spread around the average value of R, resulted from data statistics, has been taken into account. The results (95% confidence level) of the calculations are shown in the figure 13. In all cases normal contrast patterns were fitted with higher R values. This can be explained by lower signal-to-noise ratio but also by higher electron energy loss. Dynamical simulations for normal and inverted patterns were performed for the initial electron beam energies and are not accounting energy loss. Ge inverted patterns that were fitted to only 4 pixels show a low level of confidence. Both FCC Au and HCP Au show similar levels of R.

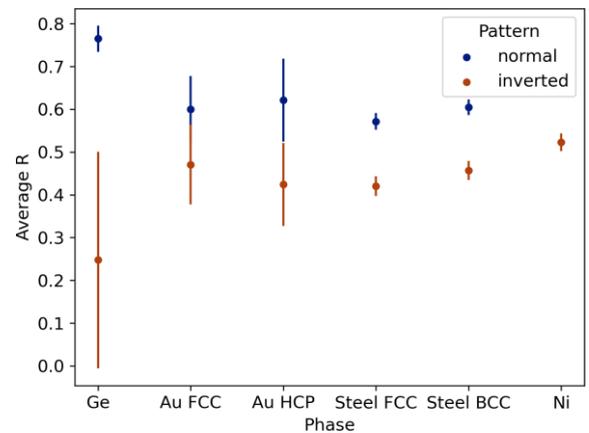

Figure 13: Average cross-correlation coefficients R for different studied phases for normal and inverted pattern matching together with calculated using software [13] error bar (95% confidence level).

## Appendix B

*Hough/Radon indexing of inverted contrast patterns*

Indexing using Hough/Radon transform was performed on Ni sample measured at 0° surface tilt. Inverted contrast experimental patterns were collected using Aztec v. 6.1. Aztec software stored inverted experimental patterns in .ebsp file format. Later pattern's contrast was inverted in .ebsp file to allow Hough/Radon indexing in Aztec software. Indexing was performed using "Optimized BD" indexing mode in Aztec v. 6.1.

The Fig. 14 Hough/Radon indexing results are presented. The indexing routine implemented in Aztec successfully indexed 94% of the points from the Ni sample. As can be seen in Fig.14 c), the software was able to detect bands correctly even though the pattern center lies below the image.

## Authors contribution

G.C Conceptualization; Investigation, Project administration; Visualization; Writing - original draft; Writing - review & editing
A.W. Software; Validation; Visualization; Writing - review & editing; Conceptualization; Funding acquisition;
G.N.Visualization, Formal analysis;Writing - original draft; Writing - review & editing; Conceptualization
T.T. Supervision; Validation

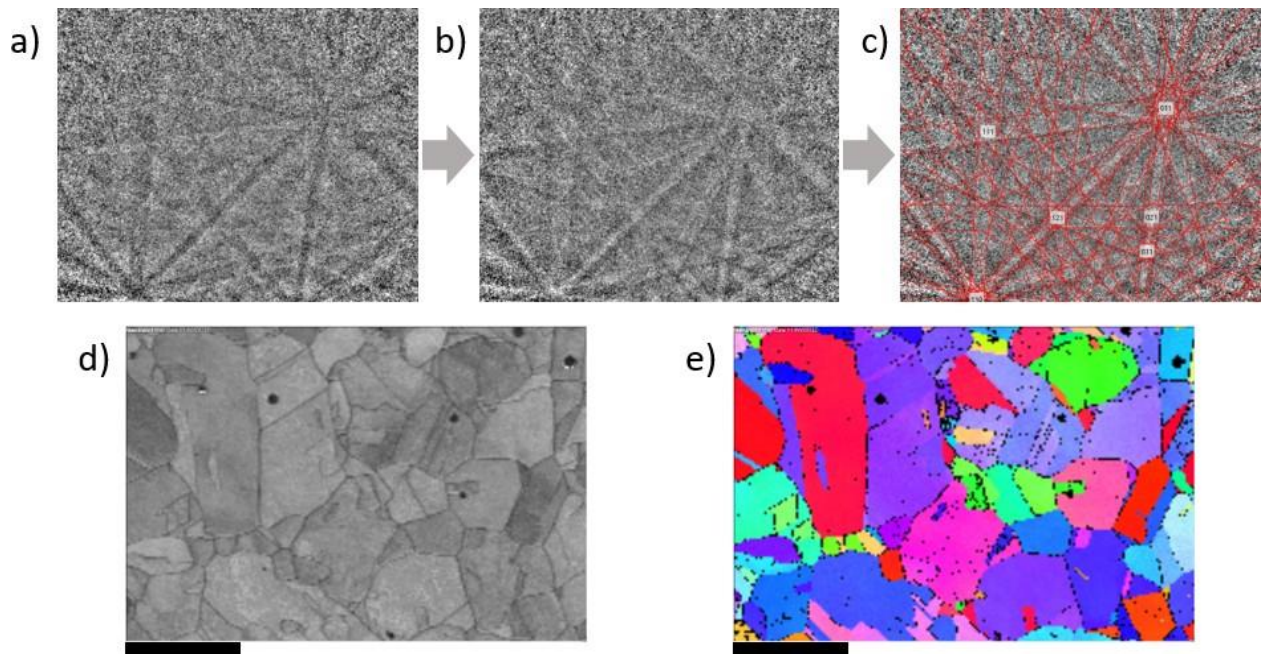

Figure 14: Inverted contrast Hough/Radon indexing of Ni sample data collected at 0°tilt, a) as collected inverted contrast pattern, b) mathematically inverted pattern from a), c) indexed b), d) band contrast, e) IPF-Z map (hit rate 94%). The scale bar is 50 microns long.

B.R.J. Formal analysis; Conceptualization
P.B. Funding acquisition; Supervision